\documentclass[conference]{IEEEtran}
\usepackage[pdftex]{graphicx}
\graphicspath{{../pdf/}{../jpeg/}}
\DeclareGraphicsExtensions{.pdf,.jpeg,.png}
\usepackage{amsmath}
\usepackage{array}
\usepackage{mdwmath}
\usepackage{cite}
\usepackage{booktabs} 
\usepackage{mdwtab}
\usepackage{eqparbox}
\usepackage{url}
\usepackage{algorithm}
\usepackage{algpseudocode}
\usepackage{physics}
\usepackage{mathtools}
\usepackage{stmaryrd} 
\usepackage{xcolor}
\usepackage{tcolorbox}
\usepackage{orcidlink}
\usepackage{amssymb}
\usepackage{hyperref}
\hypersetup{
	colorlinks=true,
	linkcolor=blue,
	filecolor=blue,      
	urlcolor=blue,
	citecolor=blue,
	pdfstartview={XYZ null null 1.07}
}
 \usepackage{subcaption}

\begin{document}

\title{\LARGE Brain Tumor Diagnosis Using Hybrid Quantum Convolutional Neural Networks} 

 \author{\authorblockN{Muhammad Al-Zafar Khan\textsuperscript{1}, Abdullah Al Omar Galib\textsuperscript{2}, Nouhaila Innan\textsuperscript{3}, Mohamed Bennai\textsuperscript{3}}
\authorblockA{\textsuperscript{1}College of Interdisciplinary Studies, Zayed University, Abu Dhabi, UAE.\\
\textsuperscript{2}Independent Researcher}
\textsuperscript{3}Quantum Physics and Spintronics Team, LPMC, Faculty of Sciences Ben M'sick, \\Hassan II University of Casablanca, Morocco\\

  \authorblockA{Muhammad.Al-ZafarKhan@zu.ac.ae, abdullahalomargalib@gmail.com, \\nouhaila.innan-etu@etu.univh2c.ma, mohamed.bennai@univh2c.ma}}
  
\maketitle

\begin{abstract}
Accurate classification of brain tumors from MRI scans is critical for effective treatment planning. This study presents a Hybrid Quantum Convolutional Neural Network (HQCNN) that integrates quantum feature-encoding circuits with depth-wise separable convolutional layers to analyze images from a publicly available brain tumor dataset. Evaluated on this dataset, the HQCNN achieved 99.16\% training accuracy and 91.47\% validation accuracy, demonstrating robust performance across varied imaging conditions. The quantum layers capture complex, non-linear relationships, while separable convolutions ensure computational efficiency. By reducing both parameter count and circuit depth, the architecture is compatible with near-term quantum hardware and resource-constrained clinical environments. These results establish a foundation for integrating quantum-enhanced models into medical-imaging workflows with minimal changes to existing software platforms. Future work will extend evaluation to multi-center cohorts, assess real-time inference on quantum simulators and hardware, and explore integration with surgical-planning systems.
\end{abstract}

\IEEEoverridecommandlockouts

\IEEEpeerreviewmaketitle
\section{Introduction}
Brain tumors are neoplasms representing abnormal cell growth within the brain and its surrounding structure. Although a relatively rare form of cancer -- in comparison to other cancers -- analogous to other cancers, it can be benign or malignant. It is poorly diagnosed because neurological symptoms cannot be detected easily, and specific tests, like biopsies and analysis of brain scans, need to be conducted by human experts. The process of diagnosis usually involves three facets: Consideration of molecular features, analysis of histological characteristics (microscopic analysis of cells and tissues from the brain), and anatomical location (lobes of the brain). 

Holistically, these tumors are categorized as follows:
\begin{enumerate}
\item \textbf{Meningiomas:} Tumors originating from the meninges surrounding the brain and the spinal cord, 
\item \textbf{Pituitary Adenomas:} Tumors originate in the pituitary gland at the base of the brain. 
\item \textbf{Gliomas:} Tumors that originate from the glial cells in the central nervous system. Gliomas are the most common type of brain tumors that are developed, and are therefore known as ``primary brain tumors.''
\item \textbf{Metastatic Lesions:} Tumors originate from cancer cells in other parts of the body, and have spread via metastatis to the brain. Therefore, they are known as ``secondary brain tumors''.
\item \textbf{Medulloblastomas:} Tumors that originate from the cerebellum, and are most commonly found in children.
\item \textbf{Acoustic Neuromas / Schwannomas:} Tumors that originate from the Schwann cells insulting nerve fibers. 
\item \textbf{Pinealomas / Pineocytoma:} Tumors that originate from the pineal gland. 
\end{enumerate}     
According to research published by the American Cancer Society \cite{ref100}, in 2023 for the state of Ohio in the United States, it was reported that $24\;810$ adults (with $14\;280$ being men, and $10\;530$ being women) in the United States were diagnosed with a form of cancerous tumors of the brain and spinal cord, and a 2020 survey found that globally $308\;102$ people were diagnosed. Although a small percentage of the overall population of the United States and the world, respectively, these patients are eligible to receive the best care possible. 

Usually, treatment paths are patient-specific, depending on several factors, such as the patient's current state of health, the presence of other underlying diseases, and so on. As with other cancers, treatment is very expensive, and requires the patient to take significant steps to a complete lifestyle overhaul. This diverse approach encompasses, amongst others, a change of diet, incorporation of exercise, radiation therapy, immunotherapy, chemotherapy, and molecular therapy. 

With the current Artificial Intelligence (AI) revolution in all fields that the world is experiencing, the medical field is no exception. Specifically, within the context of the analysis of medical images for diagnosis and prognosis, Convolutional Neural Networks (CNNs) are a type of NN architecture designed for computer vision and image processing assignments. Similar to how a perceptron is modeled after biological neurons, the CNN is modeled after the cortical preprocessing regions of the striate cortex (primary visual cortex -- V1) and the prostrate cortex (secondary visual cortex -- V2), located at the occipital lobe at the back of the brain. Analogous to how computer vision tasks in the pre-Machine Learning (ML) and pre-Deep Learning (DL) eras used to be concerned with detecting edges, shapes, and textures, the neurons in this region are sensitive to discerning these patterns. CNNs have proven to be invaluable in the medical domain, and many healthcare facilities are incorporating CNN-based classification systems together with medical experts for early disease detection, and recommended treatment plans \cite{ref101, ref200, ref201}. 

Architecturally, the CNN is described as being composed of two distinguishable layers:
\begin{enumerate}
\item \textbf{Convolutional Layer:} These contain adjustable filters/kernels/weights optimized for the task's superlative performance. The output from this layer is known as \textit{feature maps}. Given a two-dimensional input image $X$, the kernel $K$ is applied in order to get the feature map $Y$ at each spatial location
\begin{equation}
    Y(i,j)=\sum_{m}\sum_{n}X(i+m, j+n)\cdot K(m,n),
\end{equation}
where $i$ and $m$ are the $x$-coordinates of the image and kernel locations respectively, and $j$ and $n$ are the $y$-coordinates image and kernel locations respectively. Subsequently, a nonlinear activation function that serves the network in order to introduce sparsity into the network, mitigate small gradients so that the vanishing gradients problem is avoided during backpropagation, and speed up the convergence of the loss function. Typically in a CNN, a $\text{ReLU}(x)=\max(0,x)$, or a variant, is used. 
\item \textbf{Pooling Layer/Max Pooling/Subsampling:} These reduce the spatial resolution of the feature maps by a process called \textit{downsampling}, which reduces the size of the feature map, typically by half its original size. There are several types of pooling operations, these include:
\begin{enumerate}
\item \textbf{Max Pooling:} Chooses the maximum value from each subregion of the feature map.
\begin{equation}
Y(i,j)=\underset{m,n}{\max}\;X(i\times s+m, j\times s+n),
\end{equation}
where $s$ is the \textit{stride} or number of pixel shifts over the input matrix. 
\item \textbf{Average Pooling:} Chooses the average value from each subregion of the feature map.
\begin{equation}
Y(i,j)=\frac{1}{k}\sum_{m}\sum_{n}X(i\times s+m,j\times s+n).
\end{equation}
\item $\boldsymbol{\ell^{2}}$\;\textbf{Norm Pooling:} Chooses the $\ell^{2}$ norm of each subregion in the feature map.
\begin{equation}
Y(i,j)=\frac{1}{k}\sum_{m}\sum_{n}X^{2}(i\times s+m,j\times s+n).
\end{equation}
\item\textbf{Global Pooling:} Chooses the maximum or average value across all the feature maps in a layer. This produces a single scalar value for each feature map. 
\end{enumerate}
\end{enumerate}

Subsequently, once a series of convolution and pooling layers are applied (the number of each is an adjustable hyperparameter decided to maximize accuracy, and minimize time to convergence), the output is flattened, and fed to the terminal layer, where classification takes place. Generally, the sigmoid, $\sigma(x)=1/\left[1+\exp(-x)\right]$, function is used for binary classification tasks, or the softmax, $\tilde{\sigma}(\mathbf{x})=\exp(x_{i})/\sum_{j=1}^{|\text{classes}|}\exp(x_{j})$, is used, where $|\text{classes}|$ is the number of classes. 

Furthermore, the sub-operations within the network that take place include:

\textit{Subsampling} decreases the number of weights in the network, thereby increasing the receptive field of the neurons in order to capture more information about the input image on a global scale, and improve the network's ability to recognize objects, no matter their location, via \textit{translational invariance}.

\textit{Convolution} is the mathematical operation of sliding a filter matrix, containing learnable weights that help infer different features over the image, and take the inner product of the matrix with the pixel at each location to extract the feature. The importance of convolution cannot be over-stressed as it is fundamental for automatically discerning features, thus eliminating the need for manual feature engineering. However, each time a convolutional operator is applied, the dimensionality of the image is reduced.

\textit{Padding} is the process of adding pixels around the edge of an image prior to applying the convolutional operator. This process is vital because it ensures that the resulting feature map has the same dimensions as the input image. In addition, if images are not edge-pixelated, once the convolutional operator is applied.

\textit{Batch normalization} is a statistical technique used to normalize the activations of the neurons in the CNN. This is accomplished by adjusting and scaling the inputs to each layer.
\begin{figure*}[htpb]
\centering
\includegraphics[width=1\linewidth]{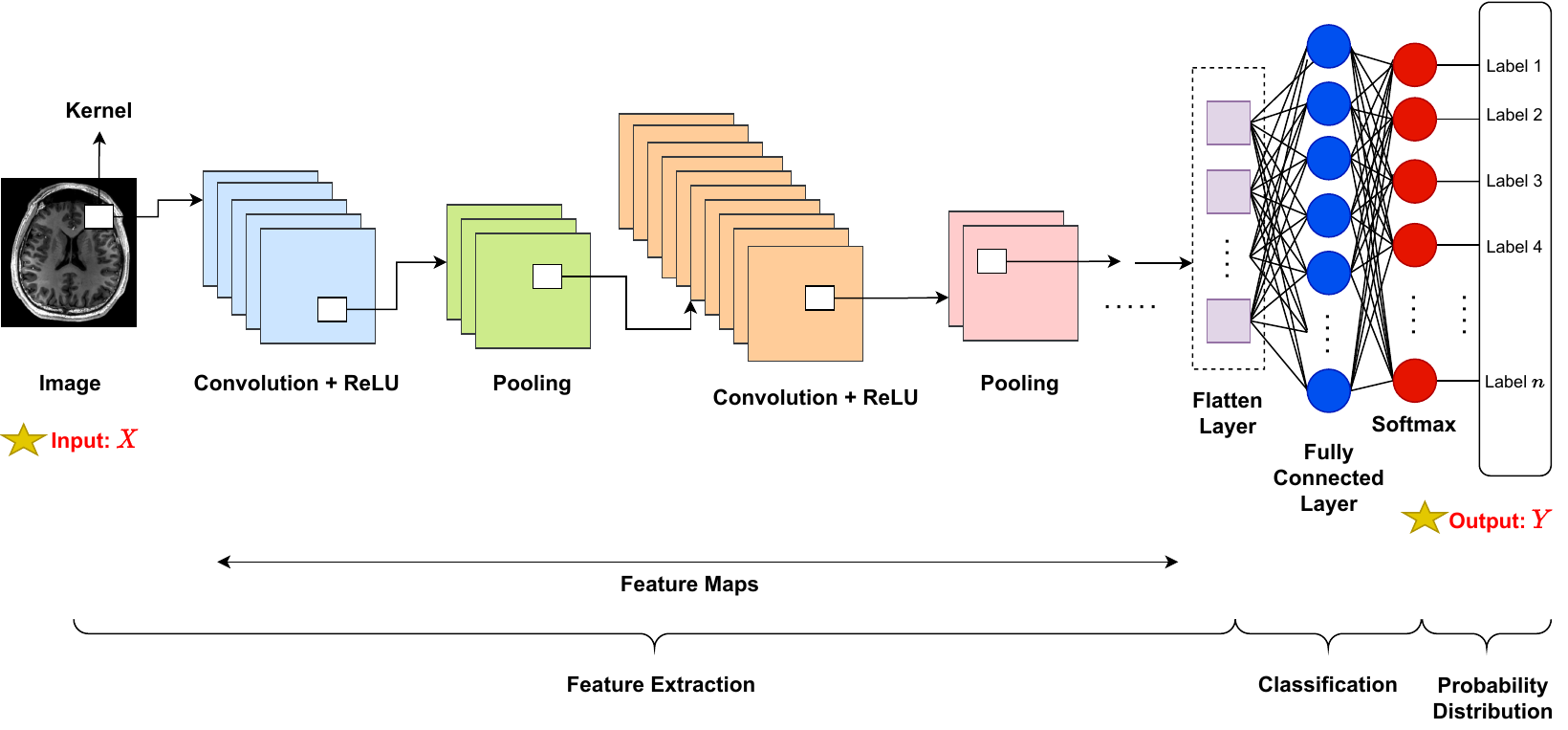}
\caption{Architecture of a classical CNN.}
\label{ccnn}
\end{figure*}

However, classical CNNs have the drawback of being unable to learn global and remote semantic information well, amongst others. Therefore, a natural consideration for advancement would be incorporating another thriving technology; thus, our proposal of supplementation via Quantum Machine Learning (QML). Below, we summarize the drawbacks of classical CNNs and highlight how QCCNs may conceivably address these challenges.

\begin{enumerate}
\item \textbf{The Problem of Overfitting:} CNNs are highly prone to learning the probability distributions of the training data ``too well'', and therefore even learn erroneous noise and outliers. \textcolor[RGB]{0,0,180}{Owing to larger feature spaces, HQCNNs are potentially less prone to overfitting}. 
\item \textbf{Cost of Computation:} Regular CNNs have quadratic runtimes, $\mathcal{O}(\mathcal{N}_{i}\times\mathcal{N}_{o}^{2}\times\mathcal{S}_{i}^{2}\times\mathcal{S}_{c}^{2})$, where $\mathcal{N}_{i}$ is the number of input feature maps, $\mathcal{N}_{o}$ is the number of output feature maps, $\mathcal{S}_{i}$ is the size of the input feature map, and $\mathcal{S}_{c}$ is the size of the convolution filter. \textcolor[RGB]{0,0,180}{HQCNNs have the potential to be much lower due to the superposition of qubits that create an exponential speedup}. 
\item \textbf{Learning Mappings from Features to the Predictor:} As the image dataset gets more complicated, and more expansive in size, CNNs limited in their expressivity. \textcolor[RGB]{0,0,180}{HQCNNs have the potential to be more expressive due to exploiting the quantum mechanical properties of superposition and entanglement}. 
\end{enumerate}

Therefore, it is evident that considering HQCNNs, and the augmentation of classical CNNs with QML proves to be a fruitful venture. Integrally, QML is a developing research enterprise, that has myriad real-world applications in diverse fields like drug discovery \cite{dd1,dd2,dd3,dd4,dd5}, materials science \cite{ms1, ms2}, condensed matter physics \cite{cmp1,cmp2, cmp3, cmp4}, optimization \cite{o1,o2,o3,o4,o5}, finance \cite{f1,f2,f3,f4,f5}, logistics planning \cite{lp1,lp2,lp3,lp4}, cryptography and cybersecurity \cite{cs1,cs2}, and many other innovative applications \cite{g1,g2,g3,g4}.   

The augmentation of QML with CNNs leads to \textit{Quantum Convolutional Neural Networks} (HQCNNs). In this study, we use HQCNNs to perform brain tumor classifications on an open-source brain tumor image dataset. While this is not novel, this study offers several contemporary advancements, and superior performance for in-distribution classification, as well as out-of-distribution generalizability.  

The original contributions of this paper are: The integration of an extremely large image dataset, which is significant considering the computational power available in the NISQ era; the development and implementation of an HQCNN with a classical component that achieves high accuracy; and a modular and scalable framework for the HQCNN that ensures adaptability to various computational contexts.

The structure of this paper is organized as follows. In Sec. \ref{sec2}, we delve into the methodologies and underlying theories of QCNNs and HQCNNs, providing a detailed theoretical background and a literature review of the most relevant papers in the field. In Sec. \ref{sec4}, we describe the process and implementation of the HQCNN used in this study, detailing the adaptations and enhancements made to the original model. In Sec. \ref{sec5}, we present the results of our experiments. In Sec. \ref{sec6}, we summarize our research findings and provide a reflective analysis of the results obtained.

\section{Background and Related Works \label{sec2}}
\subsection{Background}
\begin{figure*}[htpb]
\centering
\includegraphics[width=1\linewidth]{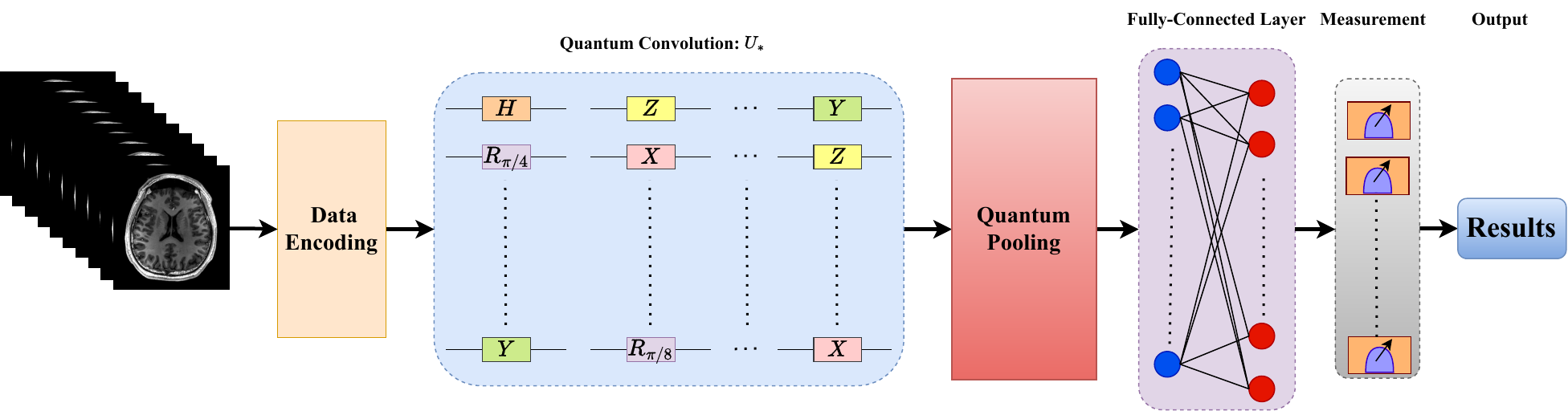}
\caption{Hypothetical architecture of QCNNs model.}
\label{qcnnarch}
\end{figure*}

Comparably similar to CNNs, QCNNs use a network of quantum convolution layers and quantum activation functions to perform feature extraction from images. Operationally, as presented in Fig. \ref{qcnnarch}, we can describe the operation of a QCNN as follows: 

\begin{enumerate}
\item \textbf{Data Encoding:} Classical data $D_{i}\in\mathcal{D}$ is converted to a single-qubit quantum state $\ket{\psi_{i}}\in\mathcal{H}$,
\begin{equation}
D_{i} \overset{\text{encoding}}{\longrightarrow} \ket{\psi_{i}} = \alpha\ket{0} + \beta\ket{1},
\end{equation}
where $\alpha,\beta\in\mathbb{C}$ are complex probability amplitudes.
\item \textbf{Quantum Convolution:} This is achieved by applying a series of quantum gates to the encoded state $\ket{\psi}$. We cumulatively represent these gates as $U_{*}$,
\begin{align}
U_{*}\ket{\psi}=U_{*}\overset{\Delta}{=}&\;\texttt{combination}(H^{\otimes n_{1}}, R^{\otimes n_{2}}_{\theta}, \nonumber \\
& X^{\otimes n_{3}}, Y^{\otimes n_{4}}, Z^{\otimes n_{5}},\ldots),
\end{align}
where the function \texttt{combination} represents taking amalgamations of the single-qubit gates for $n_{i}$, and $i=1,2,\ldots$, number of times.
\item \textbf{Quantum Pooling:} The operation is carried out by performing the \textit{quantum SWAP test}: Given two quantum states $\ket{\psi_{i}}, \ket{\psi_{j}}$ and an ancillary qubit $\ket{0}$, 
\begin{enumerate}
\item[3.1.] Apply the Hadamard gate to the ancillary qubit in order to create a superposition
\begin{equation}
H\ket{0}=\frac{1}{\sqrt{2}}\left(\ket{0}+\ket{1}\right)=\ket{+}.
\end{equation}
\item[3.2.] Apply the controlled-SWAP gate with $\ket{0}$ as the control, and $\ket{\psi_{i}}$ and $\ket{\psi_{j}}$ as the the targets,
\begin{align}
\text{CSWAP}(\ket{0};\ket{\psi_{i}},\ket{\psi_{j}}) 
    &= \ket{0}\bra{0}\otimes \ket{\psi_{i}}\bra{\psi_{i}} \nonumber \\
    &\quad + \ket{1}\bra{1}\otimes \ket{\psi_{j}}\bra{\psi_{i}}.
\end{align}
    \item[3.3] Apply the Hadamard transformation,
    \begin{equation*}
        H\ket{0}=\ket{+}.
    \end{equation*}
 \end{enumerate}
\item \textbf{Fully-Connected Layer:} This layer is composed of neurons that are oriented in a feed-forward arrangement whereby previous neurons are connected to all subsequent neurons. 
\item \textbf{Measurement:} Perform measurement to ascertain the end state. 
\end{enumerate}
In Algorithm \ref{algqcnn}, we provide a general procedure for applying the HQCNNs model to any dataset.
\begin{algorithm}
\caption{HQCNN($\mathcal{D}$)}
\begin{algorithmic}
\Procedure{HQCNN}{$\mathcal{D}$}
\State \textbf{Input}: a dataset $\mathcal{D}=\left\{D_{1},D_{2},\ldots, D_{p}\right\}$ consisting of $p$ $n\times m$-dimensional images
\State Initialize the number of repeats $q$ for the quantum convolutional and quantum pooling layers
\For{each image $D_{i}$ in $\mathcal{D}$}
    \State Encode $D_{i}$ into quantum state $\ket{\psi_{i}}$
    \Repeat
        \State Apply $U_{*}$,
        \State Measure
        \State Apply quantum SWAP test
    \Until{$q$ times}
\EndFor
\State Apply a fully connected NN to transform the output into a 1D vector of class scores
\State Measure to obtain classifications
\State \textbf{Return} classified image with associated probability
\EndProcedure
\end{algorithmic}
\label{algqcnn}
\end{algorithm}
\subsection{Related Works}
We present salient studies that have motivated our research undertaking, which have applied QML to brain tumor classification tasks, and briefly describe the results obtained. 

In \cite{ref102}, the authors developed a hybrid quantum-classical CNN (HQC-CNN) model for brain tumor classification. The novelties of the proposed network were that it achieved a high classification score (97.8\%), it was easy to train, and it converged to a solution very quickly. The classification problem was posed as a quaternary model, i.e., having four output class types: Meningioma, glioma, pituitary, and no tumor. In addition, it was demonstrated that the proposed network outperformed many well-known models. 

In \cite{ref103}, the authors use deep features extracted from the famous \texttt{Inception V3} model, combined with a parametric quantum circuit on a predictor variable with four classes, similar to the classification task in \cite{ref102}. Using three datasets as benchmarks (from Kaggle, the 2020-BRATS dataset, and locally-collected dataset), it was shown that the hybrid approach proposed here should have superior performance compared to traditional CNNs, with more than 90\% accuracy.

The research in \cite{ref104} sets out to address the ever-growing size of image datasets in medical diagnosis while maintaining patient privacy, with a specific emphasis on brain tumor images. Further, a secure encryption-decryption framework is designed to work on Magnetic Resonance Imaging (MRI) data, and a 2-qubit tumor classification model is implemented. The Dice Similarity Coefficient (DSC) was used as a validation metric, and the model was found to have a value of 98\%. This research not only considered designing a high-accuracy brain tumor classification model but also addressed the issue of cryptographic concerns. 

In \cite{ref105}, the authors aim to address concerns around tumor imaging and prediction using traditional ML techniques amongst children and teenagers. Using India as a study group, it is found that the data lacks variability, and does not account for abnormalities within these specific groups. To address this, an HQCNN model is proposed that integrates techniques from image processing to mitigate noise. It is found that the proposed quantum network achieves an 88.7\% accuracy.   

In \cite{ref106}, the authors use the MRI dataset to implement a brain tumor model using an MRI-radiomics variational Quantum Neural Network (QNN). The workflow involved using a mutual information feature selection (MIFS) technique, and the problem was converted into a combinatorial optimization task, which was subsequently solved using a quantum annealer on a D-Wave machine. While specifically not a CNN-based model, or augmentation thereof, the technique proved versatile and achieved on-par accuracies with classical methods. 

In \cite{ref107}, the authors design an automatic MRI segmentation model based on qutrits (quantum states of the form $\ket{\psi}=\alpha\ket{0}+\beta\ket{1}+\gamma\ket{2}$ with $|\alpha|^{2}+|\beta|^{2}+|\gamma|^{2}=1$), called \textit{quantum fully self-supervised neural networks} (QFS-Net) It is shown that this approach increases segmentation accuracy (model predictability) -- outperforming classical networks -- and convergence. Besides using a qutrit framework for which the model is based, the other novel contributions of this research were the implementation of parametrized Hadamard gates, the neighborhood-based topological interconnectivity amongst network layers, and the usage of nonlinear transformations. It was demonstrated that QFS-Net showed favorable results on the Berkeley gray-scale images. 

Various other literature pieces are contained within the pieces discussed above, and it will be an exercise in futility simply to repeat the findings. However, what is evident is that the hybrid approach, via augmentation of the HQCNN with salient features of the classical CNN, emphasized in the literature, serves as motivation for adopting a hybridized approach to the network in this paper. 


\section{Process and Implementation\label{sec4}}
In this section, we detail the practical steps taken to operationalize our HQCNN model by discussing the workflow. We begin by delineating the composition and sourcing of our dataset, followed by the procedures employed in data preprocessing to ensure optimal model input quality. Subsequently, we elucidate our innovative approach to quantum image processing, which sets the stage for the subsequent CNN training. This phase is critical as it underpins the model's ability to learn from the quantum-enhanced feature space. Each step is crafted to build incrementally towards a robust and practical HQCNN application.
\subsection{Dataset}
\begin{figure}[htpb]
  \centering
  \includegraphics[width=0.4\textwidth]{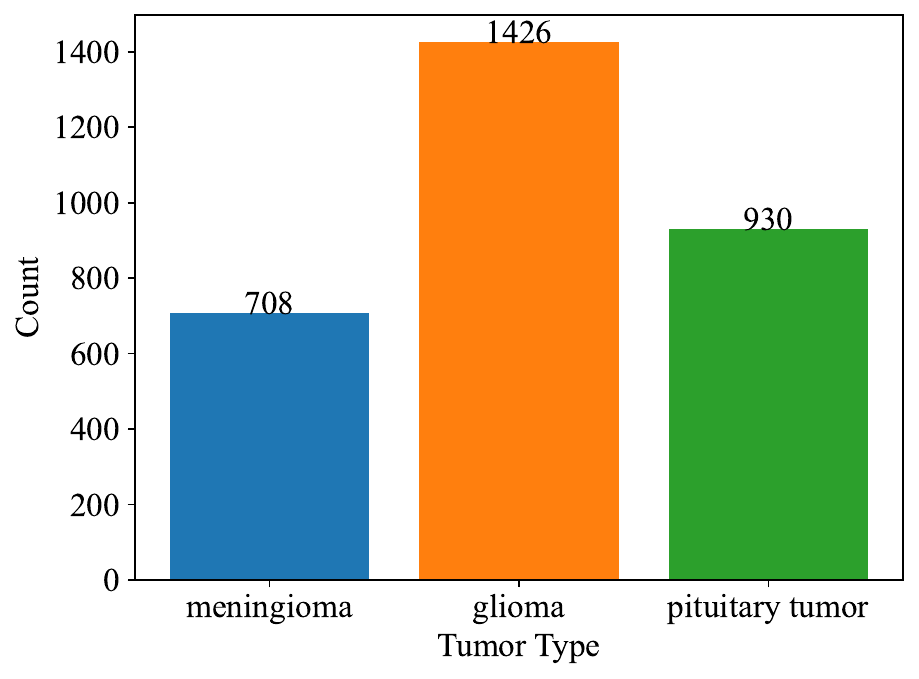}
  \caption{Distribution of sample sizes across tumor categories in the dataset.}
\label{d1}
\end{figure}

\begin{figure}[htpb]
    \centering \includegraphics[width=1\linewidth]{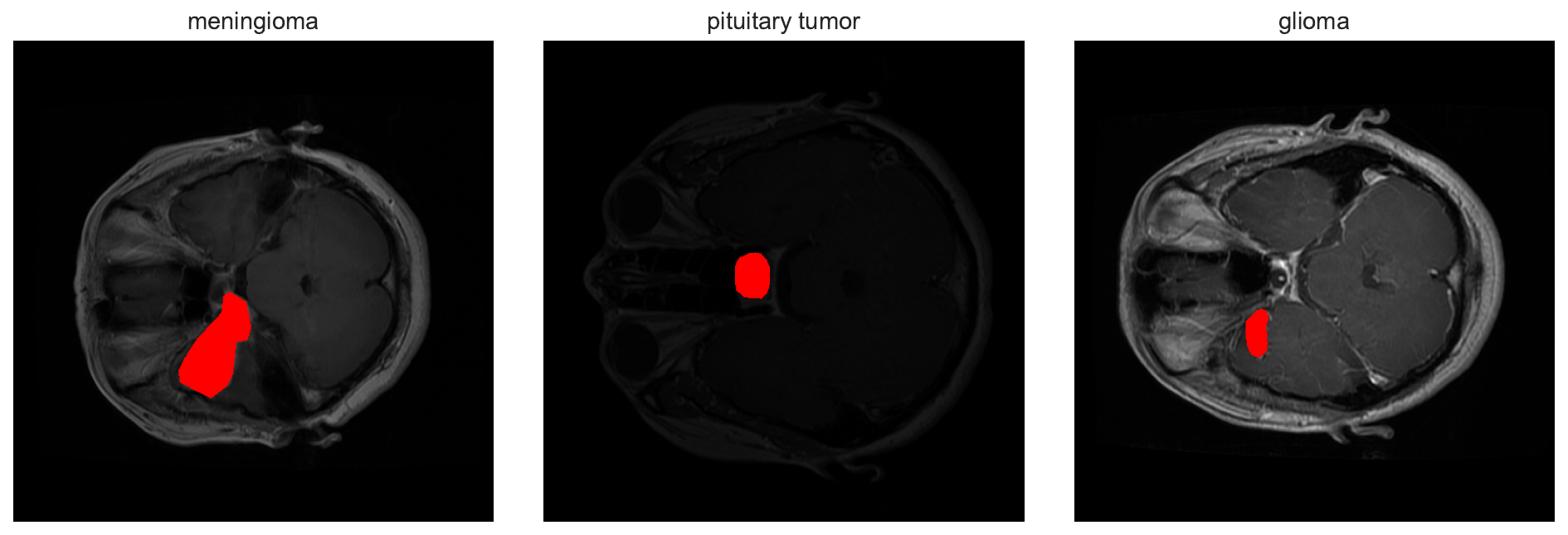}
    \caption{Diagnostic imaging examples: Brain tumor types from the dataset.}
    \label{samples}
\end{figure}

The dataset used in this research \cite{dataset} comprises $3\;064$ T1-weighted contrast-enhanced images extracted from 233 patients, encompassing three distinct brain tumor types, as shown in Fig. \ref{d1}: meningioma (708 slices), glioma ($1\;426$ slices), and pituitary tumor (930 slices). The data is organized in \texttt{MATLAB} format (.mat files), each containing a structure that encapsulates various fields detailing the image and tumor-specific information.
Each \texttt{MATLAB} (.mat) file in the dataset encompasses the following key fields:
\begin{itemize}
    \item [\textbf{a.}] \texttt{cjdata.label:} An integer indicating the tumor type, coded as follows:
    \begin{enumerate}
        \item Meningioma
        \item Glioma
        \item Pituitary Tumor
    \end{enumerate}
\end{itemize}
\begin{enumerate}
    \item[\textbf{b.}] \texttt{cjdata.PID:} Patient ID, serving as a unique identifier for each patient.
    \item[\textbf{c.}] \texttt{cjdata.image:} Image data representing the T1-weighted contrast-enhanced brain scan.
    \item[\textbf{d.}] \texttt{cjdata.tumorBorder:} A vector storing the coordinates of discrete points delineating the tumor border. The vector format is as follows:
    \begin{itemize}
        \item \parbox{3cm}{[$x_1, y_1, x_2, y_2, \ldots$]}
        \item These coordinate pairs represent planar positions on the tumor border and were generated through manual delineation, offering the potential to create a binary image of the tumor mask.
    \end{itemize}
    \item[\textbf{e.}] \texttt{cjdata.tumorMask:} A binary image where pixels with a value of 1 signify the tumor region. This mask is a crucial resource for further segmentation and analysis.
\end{enumerate}
This dataset provides a diverse set of brain scans, as shown in Fig. \ref{samples}, and includes manual annotations of tumor borders, fostering the development and evaluation of robust image processing and ML models for brain cancer classification \cite{d2,d22}.
\subsection{Data Preprocessing}
This step loads the MATLAB data files using h5py and extracts image data, tumor labels, and relevant information. The T1-weighted images (originally 512x512 pixels) are resized to 25\% of their original dimensions (approximately 128x128 pixels) and normalized to a [0,1] range to ensure compatibility with quantum circuit processing and subsequent neural network training.

\subsection{Quantum Image Processing}
\begin{figure*}[htpb]
    \centering    
    \includegraphics[width=1\linewidth]{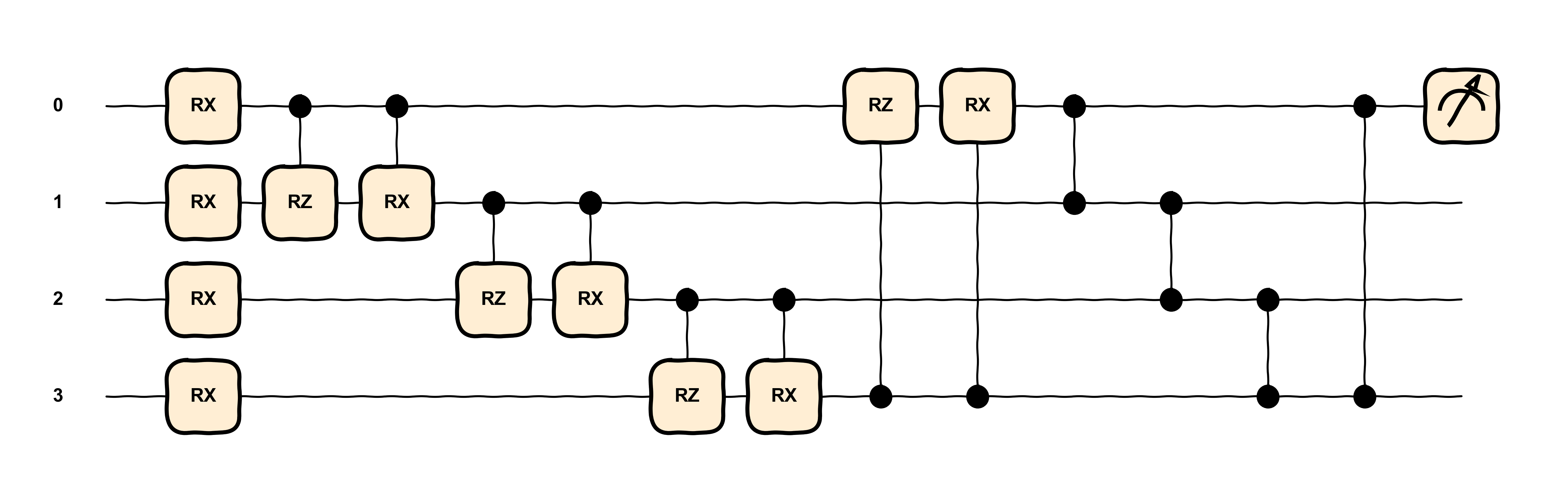}
    \caption{HQCNN circuit.}
    \label{qcnnc}
\end{figure*}

In this step, the HQCNN circuit stands as a pivotal component, and the circuit comprises several essential elements. Initially, a quantum device with four qubits is initialized using \texttt{PennyLane's} default qubit simulator \cite{pennylane}, as presented in Fig.~\ref{qcnnc}. A significant step in this process is setting the parameter $\theta$ to $\pi/2$. This parameter finds its application in Controlled-Rotation-$Z$ ($\text{CR}_{Z}$) and Controlled-Rotation-$X$ ($\text{CR}_{X}$) gates.
The circuit employs Rotation-X ($\text{R}_X$) gates applied to each qubit. These gates rotate the qubit around the $X$-axis of the Bloch sphere. The rotation angle is proportional to the pixel value times $\pi$. This operation is mathematically expressed as 
\begin{equation}
\ket{\psi'} = \bigotimes_{i = 0}^{3} \text{R}_X(\phi_i\pi) \ket{\psi},
\end{equation}
with
\begin{equation}
\text{R}_{X}\left(\phi_i \pi\right) =  
\begin{pmatrix}
 \cos\phi_i \pi/{2}& - \imath \sin\phi_i \pi/2\\ 
 - \imath \sin\phi_i\pi/2 & \cos\phi_i \pi/2
\end{pmatrix},
\end{equation}
where \(\ket{\psi} = \ket{0}^{\otimes 4}\) represents the initial state of the quantum system, and \(\phi_i\) are the elements of the pixel value array passed to the quantum circuit.

Following this, the circuit incorporates controlled rotation gates, specifically $\text{CR}_{Z}$ and $\text{CR}_{X}$ gates. These gates are applied between each pair of adjacent qubits and between the last and first qubits. These gates introduce interactions between the qubits. Mathematically, this is represented as 
\begin{align}
\ket{\psi''}=&\;\left(\text{CR}_{Z}(\theta)_{0,1} \cdot \text{CR}_{X}(\theta)_{0,1}\right)\nonumber \\ 
&\times (\text{CR}_{Z}(\theta)_{1,2} \cdot \text{CR}_{X}(\theta)_{1,2}) \nonumber \\
&\times  (\text{CR}_{Z}(\theta)_{2,3} \cdot \text{CR}_{X}(\theta)_{2,3}) \ket{\psi'},
\end{align}
with
\begin{equation}
\text{CR}_{Z}(\theta) = 
\begin{pmatrix}
 1 &  0 & 0  & 0 \\ 
 0 & 1 & 0 & 0 \\ 
 0 & 0  & \exp(-\imath \theta/2) & 0\\ 
 0 & 0 & 0 & \exp(\imath \theta/2)
\end{pmatrix},
\end{equation}
and
\begin{equation}
\text{CR}_{X}(\theta) = 
\begin{pmatrix}
 1 &  0 & 0  & 0 \\ 
 0 & 1 & 0 & 0 \\ 
 0 & 0  &\cos\theta/2 &-\imath\sin\theta/2\\ 
 0 & 0 & -\imath\sin\theta/2 &\cos\theta/2 
\end{pmatrix},
\end{equation}
where $\text{CR}_{Z}(\theta)$ and $\text{CR}_{X}(\theta)$ are the phase-parameterized $\text{CR}_{Z}$ and $\text{CR}_{X}$ gates, respectively.
Moreover, the circuit incorporates CZ gates that are applied between each pair of adjacent qubits, and between the last and first qubits, we add an additional layer of entanglement between the qubits; mathematically, this is represented as
\begin{equation}
\ket{\psi'''} = \text{CZ}_{0,1} \cdot \text{CZ}_{1,2} \cdot \text{CZ}_{2,3} \cdot \text{CZ}_{3,0} \ket{\psi''},
\end{equation}
with
\begin{equation}
\text{CZ} = 
\begin{pmatrix}
1 & 0 & 0 &  0 \\ 
 0 & 1  & 0  & 0 \\ 
 0 & 0  & 1  & 0 \\ 
 0 & 0  & 0  & -1 
\end{pmatrix},
\end{equation}
Finally, the measurement phase occurs. In this phase, the expectation value of the Pauli-$Z$ operator is measured on the first qubit. This step provides the average result of numerous measurements, mathematically encapsulated as 
\begin{equation}
\mathcal{M} = \left\langle\phi^{\prime\prime\prime\prime\prime}\left|Z\right|\phi^{\prime\prime\prime\prime\prime}\right\rangle,
\end{equation}
where $\ket{\phi^{\prime\prime\prime\prime\prime}}$ is the state after the CZ gates, indicating the output of the quantum convolution operation on the $2\times2$ patch as a complex function of the pixel values, reflective of the quantum nature of the operation. 

The quantum convolution function plays a crucial role in the process of quantum image processing. This function is designed to perform a quantum convolution operation on an input image. The function requires two arguments: Firstly, an image represented as a 2D \texttt{NumPy} array, where each element corresponds to a pixel value. Secondly, the step size, which divides the image into patches, is set to 2 by default. Thus, the image is divided into $2\times2$ patches.

In terms of processing, the function begins by initializing a new 2D array of zeros. The dimensions of this array are determined by the height and width of the image divided by the step size, and this array serves to store the output of the quantum convolution operation. Subsequently, the function divides the image into $2\times2$ patches by looping over the image with the specified step size. For each $2\times2$ patch, the function flattens the patch into a 1D array and then applies the quantum convolution operation, as performed by the HQCNN circuit. The measurement result of this operation is then stored in the corresponding position in the output array.

In the output layer, the function returns a new 2D array. Each element in the array represents the result of the quantum convolution operation on the corresponding $2\times2$ patch in the original image. This resulting array can be perceived as a transformed version of the original image, where the transformation is a complex function of the pixel values stemming from the quantum nature of the operation.

The core component of the implementation is a loop designed for image processing, which iterates through every file in a designated folder, provided that the current file has been processed during each iteration. If it has not, it loads the image and its associated label from the file. Following this, it resizes and normalizes the image. Subsequently, a quantum convolution function is applied to the image. Finally, the processed image and its label are saved into a new file.
\subsection{CNN Training}
The approach trains a CNN on quantum-processed image data using an enhanced architecture. The model comprises a Flatten layer, followed by BatchNormalization to stabilize training. It includes Dense layers with 256 and 128 units (using He-uniform initialization), LeakyReLU activation (alpha=0.1), BatchNormalization, and Dropout (0.3 and 0.2 rates) to prevent overfitting. The output layer is a Dense layer with 4 units and softmax activation for classification into three tumor types plus a potential no-tumor class. The model is optimized using Adam with an exponentially decaying learning rate (initial rate 0.001, decay steps 1000, decay rate 0.9), and trained with sparse categorical crossentropy loss.


\section{Results and Discussion\label{sec5}}
This section presents a comprehensive analysis of the HQCNN's performance on the brain tumor classification task. All experiments were conducted over 16 training epochs using a 70/30 train/validation split, implemented using Pennylane and TensorFlow~\cite{ref102,ref101}. The training and testing were performed on a 64-bit system with an 11th Gen Intel(R) Core(TM) i5-1135G7 CPU @ 2.40GHz and 32.0 GB RAM. Full hyperparameter details and code are available in our public repository\footnote{\url{https://github.com/ahkatlio/QCNN_for_brain_cancer}}.
\begin{figure*}[htpb]
    \centering
    \begin{subfigure}[0.5]{0.45\linewidth}
        \includegraphics[width=\linewidth]{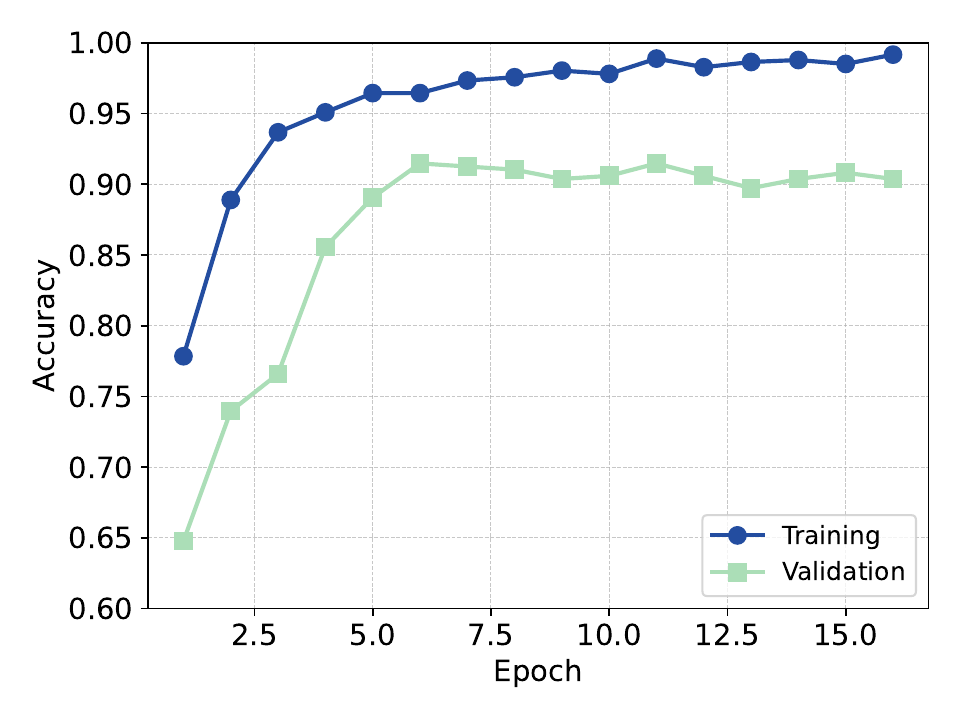}
        \caption{}
        \label{acc}
    \end{subfigure}
    \hfill
    \begin{subfigure}[0.5]{0.45\linewidth}

                \includegraphics[width=\linewidth]{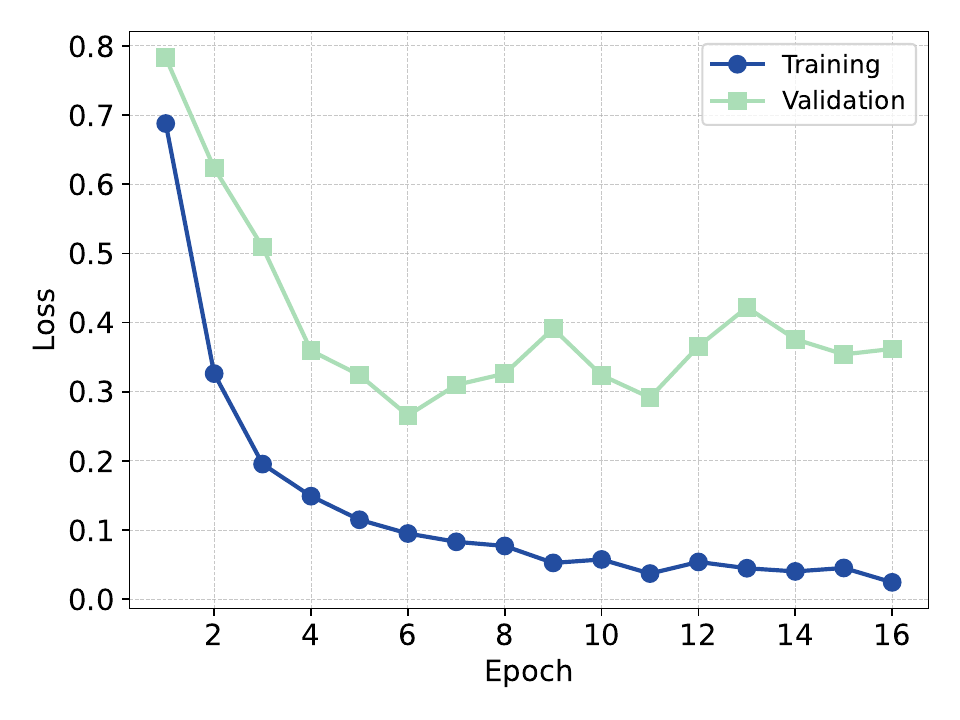}
        \caption{}
        \label{loss}
    \end{subfigure}
    \caption{(a) Training and validation accuracy per epoch, showing rapid convergence—both curves plateau by the sixth epoch with final values of approximately 99\% (training) and 91\% (validation).
(b) Training and validation loss per epoch, illustrating a steep decline in training loss to below 0.05 by epoch 11 and modest validation-loss fluctuations around 0.30–0.40, indicative of stable generalization.}
    \label{loss_and_acc}
\end{figure*}
\subsection{Convergence Behavior}

Fig. \ref{acc} illustrates the epoch‐wise accuracy for both training and validation. The training curve rises sharply, surpassing 95\% by epoch 4 and plateauing at 99.16\% by epoch 6. The validation curve follows a similar trend, reaching 90\% by epoch 5 and stabilizing at 91.47\% by epoch 6, with minimal subsequent fluctuations. Fig.~\ref{loss} shows the corresponding loss curves: the training loss decreases rapidly from 0.69 to 0.05 within the first 11 epochs, while the validation loss falls to approximately 0.30 by epoch 6 and remains within 0.30–0.40 thereafter. The close alignment of training and validation trajectories indicates that the HQCNN avoids significant overfitting and learns feature representations effectively.

\subsection{Class‐wise Performance}
\begin{figure}[htpb]
    \centering
    \includegraphics[width=1\linewidth]{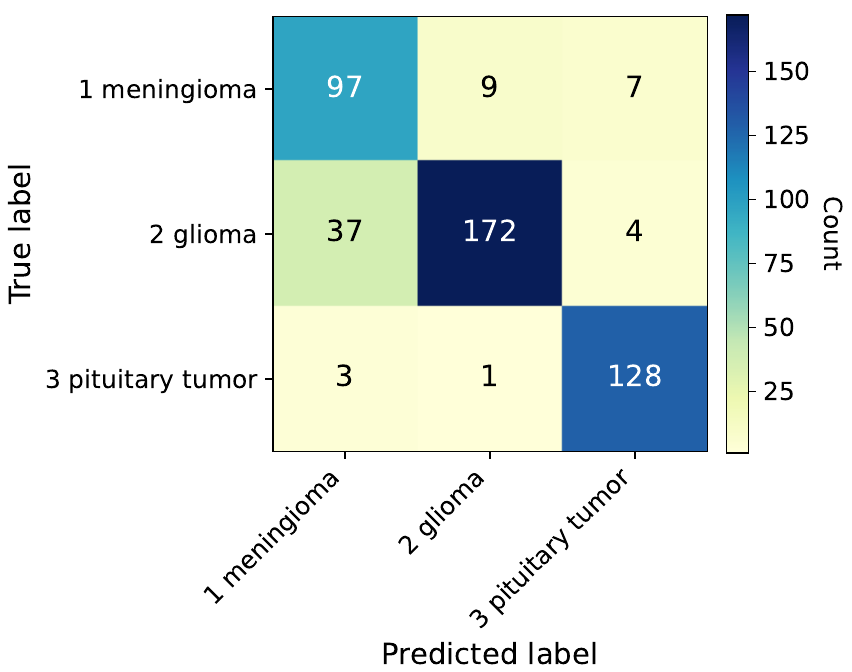}
    \caption{Confusion matrix on the test set. Pituitary tumors are classified with near‐perfect precision and recall. Most errors occur between meningioma and glioma, reflecting their visual similarity.}
    \label{cm}
\end{figure}
The confusion matrix in Fig.~\ref{cm} details class‐specific predictive performance on the held‐out test set. All three tumor classes—meningioma, glioma, and pituitary tumor—demonstrate strong performance, with recalls of approximately 85.8\%, 80.8\%, and 97.0\%, respectively. Pituitary tumors are classified with near‐perfect precision and recall, confirming the model's robustness in recognizing this class. In contrast, the primary misclassifications occur between meningioma and glioma, where the model shows reduced precision for meningioma (70.8\%) due to frequent confusion with glioma. These errors likely stem from overlapping radiographic features between these two tumor types. While the HQCNN architecture effectively captures discriminative patterns, especially for pituitary tumors, targeted enhancements such as domain-specific data augmentation or refined feature extractors may further reduce inter-class confusion, particularly between glioma and meningioma.

\subsection{Qualitative Evaluation of Model Predictions}
\begin{figure*}[htpb]
    \centering
    \includegraphics[width=1\linewidth]{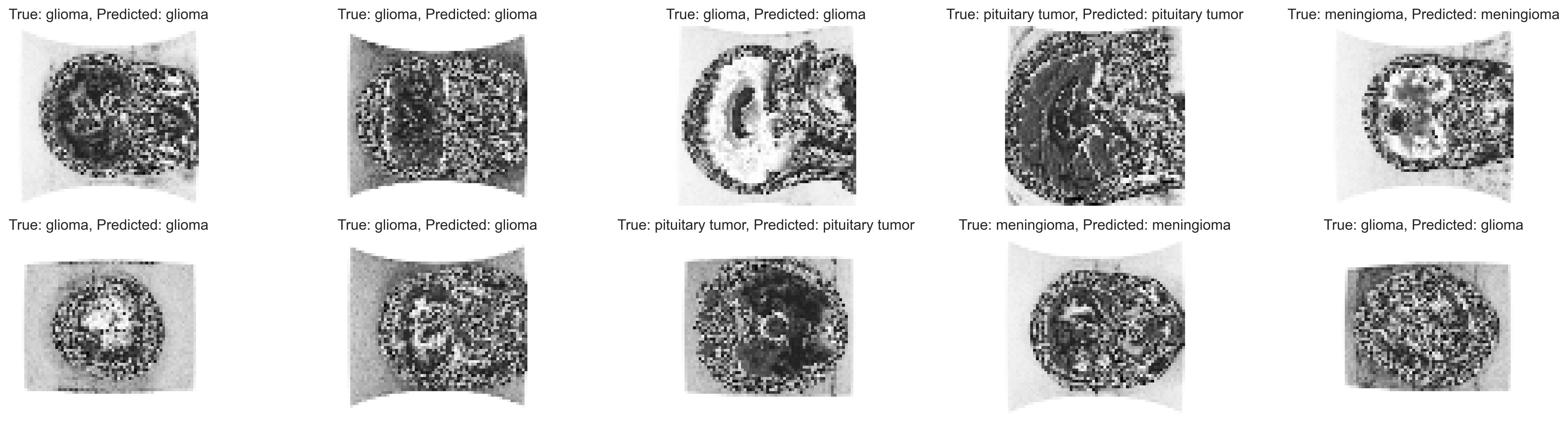}
    \caption{Randomly selected test‐set samples annotated with true and predicted class labels. All ten samples are correctly classified by the HQCNN, reflecting the model's strong generalization to unseen data.}    \label{4}
\end{figure*}
Beyond aggregate performance metrics, qualitative assessment offers valuable insights into the model's behavior on individual cases. Fig.~\ref{4} presents a representative sample of ten test‐set images selected at random, each annotated with the corresponding ground truth and the HQCNN's predicted label. The model correctly classifies all ten samples, demonstrating its ability to extract discriminative features across tumor types in real-world, heterogeneous imaging conditions. These visual confirmations align with the trends observed in the confusion matrix (Fig.~\ref{cm}) and further support the model's reliability for practical diagnostic use. Although such examples do not replace formal statistical validation, they provide intuitive evidence of the model's robustness on unseen data.

\subsection{Comparison to Existing Models}
\begin{table}[htpb]
\centering
\begin{tabular}{llll}
\toprule
\textbf{Ref} & \textbf{Model} & \textbf{Dataset} & \textbf{Accuracy} \\ \midrule
\cite{paper1} & QVC & Local Hospital & 90\% \\ 
\cite{paper2} & HQ-C CNN & Kaggle & 97.8\% \\ 
\cite{paper3} & QCNN & Public Brain Tumor& 88.7\% \\ 
\cite{paper4} & AHQCNN & REM-BRANDT & 98.07\% \\ 
-- & \textbf{Our HQCNN} & Brain Tumor & \textbf{91.47\%} \\ \bottomrule
\end{tabular}
\caption{Comparison of QML models on different brain tumor datasets.}
\label{table:brain_tumor_detection_models}
\end{table}
Table~\ref{table:brain_tumor_detection_models} compares pure quantum‐circuit CNN architectures. Our HQCNN achieves a 91.47\% test accuracy—exceeding QVC (90.0\%) and QCNN (88.7\%), though somewhat below the AHQCNN benchmark (98.07\%). This result confirms that our hybrid design delivers competitive performance among quantum models while maintaining a lightweight circuit depth that may facilitate near‐term implementation.

\subsection{Discussion}

The HQCNN model demonstrated strong performance on the brain tumor classification task, with rapid convergence and consistent accuracy across training and validation sets. Its ability to generalize from limited training epochs reflects an effective integration of classical and quantum components. The low variance between training and validation metrics suggests that the model is not overfitting despite its expressive capacity.
Performance across tumor types revealed notable strengths and limitations. Pituitary tumors were classified with high precision and recall, while the primary challenge lay in differentiating meningioma from glioma, two classes with known visual similarities in MR images. This inter-class confusion, also observed in prior works, highlights the need for domain-specific enhancements, such as data augmentation targeting subtle texture differences or attention-based feature refinement.

In comparison to other quantum CNN models, the HQCNN achieves competitive accuracy while maintaining architectural simplicity. Although models like AHQCNN report higher accuracy on different datasets, our HQCNN offers a favorable balance between model complexity and performance, making it more practical for deployment, especially in settings where quantum resources are constrained.
Furthermore, the qualitative inspection of randomly selected test cases confirmed the model's stability, with all samples correctly classified. This visual confirmation aligns with the statistical analysis and reinforces the model's robustness. However, future studies should incorporate more extensive evaluation protocols, including uncertainty estimation and model explainability tools, to better understand decision boundaries and improve trust in clinical use.

\section{Conclusion \label{sec6}}
This work introduces an HQCNN tailored for brain tumor classification using MRI data. The model achieves a test accuracy of 91.47\% and exhibits strong class-wise performance, particularly in distinguishing pituitary tumors. Its training efficiency and minimal overfitting confirm the viability of combining quantum circuits with conventional deep learning for medical imaging tasks.
While performance remains competitive with existing quantum models, further improvements are needed to reduce confusion between visually similar tumor types. Enhancing feature discrimination through tailored augmentation, interpretability techniques, and integration of multi-sequence imaging data represents a promising path forward.
In summary, the HQCNN architecture offers a practical step toward deploying quantum-enhanced neural networks in clinical diagnostics. Its simplicity, accuracy, and compatibility with near-term quantum devices highlight its potential for real-world applications in precision medicine.


\end{document}